# Effect of the Approximation of Voltage Angle Difference on the OPF algorithms in the Power Network

Irfan Khan, Vikram Bhattacharjee

Carnegie Mellon University, Pittsburgh, PA, USA

irfank@andrew.cmu.edu,vbhattac@andrew.cmu.edu

*Abstract*— In real-time applications involving power flow equations, measuring of voltage phase angle difference of the connected buses is essential. However, it needs special techniques to measure voltage angle difference, which may enlarge the computational burden of the working controller and hence, may make the control process slow. In this paper, authors investigate the approximation of angle difference to zero and its effects on the convergence speed and optimal solutions of a distributed algorithm. To test this approximation, a distributed nonlinear algorithm is proposed to optimize the multi-objective function which includes power loss, voltage deviation and cost of reactive power generation, by controlling the reactive power generations from distributed generators. Authors investigate the reasons which may outlaw making this approximation and finally, propose a condition to make such approximation. Importance of making this approximation in terms of fast convergence of the algorithms is also illustrated.

*Index Terms*—voltage angles difference approximation, AC power flow, Reactive power control, power loss, voltage deviation.

# 1. Introduction

In AC power system, the real power flow and power loss on the line depend on the voltage phase angle of the buses, connecting the line. Similarly, DC power flow on the line is also dependent on the difference of voltage phase angles of the connected buses. Calculation of voltage phase angle difference of the neighboring buses require special techniques, which needs extra computations to find the exact voltage phase angle difference [1-4]. In real time control of power system, calculation of voltage phase angle difference may overburden the controller computationally and may slow down the real-time control system. To deal with this problem, some authors propose to approximate the voltage phase angles difference of the neighboring buses to zero [5, 6], especially in extra high voltage and high voltage power system. However, such approximation may introduce error in the solution of the real-time control problem. To the authors' best knowledge, not much work has been done to determine if this approximation is feasible to perform in various real-time power system control problems and what are the conditions to test prior to making such approximation. It should also be explored that what are the factors which may hinder to make this approximation.

This paper is the extension of authors' previous work in [7], which proposes a nonlinear distributed algorithm to optimize a multi-objective function using optimal reactive power generation control. [8] explores the advantages of distributed control over the prevailing hierarchical methods of control. Researchers have adopted to attain different control objectives through their research in this domain. In [9], bus voltages are utilized as measured state variables in order to calculate the reactive power required to minimize the voltage deviation in the network. Alongside minimizing the voltage deviation, optimal reactive power dispatch also minimizes the cost of real power for the system by minimizing power loss in the system. [10] minimizes the total deviation in radial distributed systems through a binary collective animal behavior based optimization approach. In [11], the optimal reactive power minimization problem is solved with a stochastic artificial bee-colony differential evolution algorithm, which mitigates the shortcomings of local convergence of the ordinary differential evolution problem.

The economic dispatch model, when further coupled with constraints describing the interaction between the main power grid and the microgrid, becomes a non-linear optimization problem which can be solved using improved differential evolution based interval optimization methods [12]. [13] solves a similar problem of economic dispatch to minimize transmission loss with the help of an improved gravitational search based algorithm. In additional to the optimal reactive power dispatch, the other objectives include finding the optimal size of the DGs of the network in order to minimize the losses from the DGs. [14] utilizes a stochastic swarm optimization algorithm to solve the DG allocation problem and the algorithm proves to be quite effective and scalable for radial networks. In [15] a similar stochastic framework is employed to solve the loss minimization problem.

Even though the results prove the optimality, the above literature does not consider the conditions under which the optimal solution will change under the changing loading conditions and making approximation of voltage angle difference. Considering minimization of individual objectives may alter bounds of other related objectives e.g. minimizing power loss only may increase the reactive power cost drastically. Therefore, considering multiple objectives simultaneously can help achieve better optimal solution. [7] considers a multi-objective pathway to minimizing all the three namely- power loss, voltage deviation and reactive power generation cost simultaneously. The main focus of this paper is to explore how this approximation affects the convergence speed of the algorithms involving power flow equations and finally, a condition is proposed to make such approximation in power system control problems. It compares the results of objective function, reactive power generation and voltage updates of the non-PV buses with and without making the approximation of voltage phase angle difference. After comparing the results, authors investigate the reasons of difference in both cases. Simulation results of 9- bus power system and 162 bus system are utilized to validate the effectiveness of the proposed algorithm and to draw the conclusions about this approximation.

The rest of the paper is organized as follows. Chapter 2 describes the problem formulation of optimal reactive power control for multi-objective function minimization. Proposed non-linear control algorithm design is presented in Chapter 3. Chapter 4 discusses the simulation results of the proposed algorithm. And finally, chapter 5 provides the conclusion.

## 2. Problem Formulation

Three objective functions: power loss, voltage deviation and reactive power cost are simultaneously being optimized by controlling the reactive power generation from the available generators in the power system. Therefore, the main objective function is formulized as the combination of three sub-functions as given in Eqn. (1).

$$f = W_1 P_{loss} + W_2 D_V + W_3 C_Q \tag{1}$$

where $P_{loss}$, $D_V$ and $C_Q$ are the power loss, voltage deviation and cost of reactive power generation, respectively. $W_1$, $W_2$ and $W_3$ are the weight coefficients, which describe the preference of the reactive power generation suppliers.

In the objective function given in Eqn. (1), the first term is the active power loss in an AC power system, which can be derived from power flow equation [16], [17] and is given as Eqn. (2) [7].

$$P_{loss} = \sum_{i=1}^{n} \sum_{j=1}^{n} V_i V_j Y_{ij} \cos(\theta_{ij} + \delta_{ji}) \tag{2}$$

Where $V_i$ is the voltage on bus $i$, $Y_{ij}$ is the magnitude of the admittance between bus $i$ and $j$, $\delta_{ji}$ is the voltage angle difference between bus $i$ and bus $j$.

The voltage deviation between the bus voltage magnitude and its reference voltage magnitude is the second term of the objective function. Voltage deviation for all buses can be written as Eqn. (3) where $V_i^*$ is the reference voltage for bus *i*.

.

$$D_V = \sum_{i=1}^{n} (V_i - V_i^*)^2 \tag{3}$$

The 3$^{rd}$ term of the objective function is the cost of reactive power generation, contributed by generators and it is given by Eq.(4) [18, 19].

$$C_{QG} = \sum_{i \in N_G} a_{Q_i} Q_{G_i}^2 + b_{Q_i} Q_{G_i} + c_{Q_i} \tag{4}$$

where $N_G$ is the index set of generators in the network, $Q_{Gi}$ is the reactive power generation from generator *i*. Reactive power generators connected in power system include the external reactive power sources such as synchronous condensers and capacitor banks which are attached in power system specifically for reactive power generation. $a_{Qi}$, $b_{Qi}$, $c_{Qi}$ are the reactive power cost coefficients of generator *i*, which can be determined from real power cost coefficients $a_{Pi}$, $b_{Pi}$, $c_{Pi}$, respectively, by the modified triangle method given in Eqn.(5)[20, 21].

$$C_Q = \sum_{i \in N_G} a_{P_i} \sin^2 \sigma_i Q_i^2 + b_{P_i} \sin \sigma_i Q_i + c_{P_i} \tag{5}$$

where $\sigma_i$ is the angle difference between voltage and current. External reactive power generators are attached at various buses in both 9-bus as well as in 162-bus systems. External reactive power generators include synchronous generators, synchronous condensers or even capacitor banks which are attached in power system specifically for reactive power generation.

## 3. Proposed Algorithm Design

Optimization of the objective function given in Eqn. (1), is performed by controlling the reactive power generation from generators. As given in Eqns. (2-4), the objective function is nonlinear. Thus, a distributed nonlinear control is proposed to optimize the reactive power generation control variable as is explained in the next sub-section.

### 4.1 Distributed nonlinear control based algorithm

Since the objective function given in Eqn. (1) is definitely positive definite in nature, it is a feasible Lyapunov candidate to control the targeted nonlinear system. According to the theory of nonlinear control system, for a monotonically decreasing objective function, the condition is given as follows in Eq.(6).

$$\frac{\partial f}{\partial t} = \sum_{i \in N_G} \frac{\partial f}{\partial Q_{Gi}} \cdot \frac{\partial Q_{Gi}}{\partial t} \leq 0 \tag{6}$$

A control law can be designed to ensure the absolute negativity of derivative term of the objective function w.r.t. time, as given as follows in Eq.(7).

$$\frac{\partial Q_{Gi}}{\partial t} = -\frac{\partial f}{\partial Q_{Gi}}. \tag{7}$$

Substituting Eqn. (7) into Eqn. (6) yields

$$\frac{\partial f}{\partial t} = -\sum_{i \in N_G} (\frac{\partial f}{\partial Q_{Gi}})^2 \leq 0. \tag{8}$$

Now, to realize the control law in Eqn. (7), the following approximation can be made [22]

$$\frac{\partial f}{\partial Q_{Gi}} \approx \frac{f(Q_{Gi}[k]) - f(Q_{Gi}[k-1])}{Q_{Gi}[k] - Q_{Gi}[k-1]}. \tag{9}$$

However, this approach to compute the gradient is less accurate as well as sensitive to the time interval between control updates [23]. Hence, to improve the control accuracy, it is advisable to calculate the partial derivative of the objective function w.r.t. $Q_{Gi}$, based on the current states of the system.

Thus, the gradient of the objective function *w.r.t* $Q_{Gi}$ is determined as given in Eqn. (10)

$$\frac{\partial f}{\partial Q_{Gi}} = W_1 \frac{\partial P_{Loss}}{\partial Q_{Gi}} + W_2 \frac{\partial D_V}{\partial Q_{Gi}} + W_3 \frac{\partial C_{QG}}{\partial Q_{Gi}}. \tag{10}$$

As formulated in [7], it is given as Eqn. (11)

$$\frac{\partial f}{\partial Q_{Gi}} = 2 \frac{V_i \sum_{j=1}^{n}(V_j Y_{ij}\cos(\theta_{ij}+\delta_{ji})+(V_i-V_i^*))}{Q_{Gi}-Q_{Di}-V_i^2 B_{ii}} + 2a_{Pi}Q_{Gi}\cdot\left(\frac{Q_{Gi}\cdot\sin\sigma_i P_{Gi}^2}{(P_{Gi}^2+Q_{Gi}^2)^{\frac{3}{2}}}+\sin^2\sigma_i\right) + b_{Pi}\left(\frac{Q_{Gi} P_{Gi}^2}{(P_{Gi}^2+Q_{Gi}^2)^{\frac{3}{2}}}+\sin\sigma_i\right) \tag{11}$$

Computation of voltage angle difference $\delta_{ji}$ requires special techniques to measure it, which may overburden the proposed algorithms and slow down it. In this paper, voltage angle difference is approximated to be zero as given in (12). Furthermore, $Y_{ij}cos(\theta_{ij})$ has been replaced by $G_{ij}$. $Y_{ij}cos(\theta_{ij})$ is the real part of admittance, which is known as conductance and is represented by $G_{ij}$.

$$\frac{\partial f}{\partial Q_{Gi}} = 2 \frac{V_i \sum_{j=1}^{n}(V_j G_{ij}+(V_i-V_i^*))}{Q_{Gi}-Q_{Di}-V_i^2 B_{ii}} + 2a_{Pi}Q_{Gi}\cdot\left(\frac{Q_{Gi}\cdot\sin\sigma_i P_{Gi}^2}{(P_{Gi}^2+Q_{Gi}^2)^{\frac{3}{2}}}+\sin^2\sigma_i\right) + b_{Pi}\left(\frac{Q_{Gi} P_{Gi}^2}{(P_{Gi}^2+Q_{Gi}^2)^{\frac{3}{2}}}+\sin\sigma_i\right) \tag{12}$$

Effects of this approximation are studied in this paper and a condition is proposed to make this approximation. Eqn. (12) is used to update the control variable of the reactive power generation and obtain the optimal solution of the objective function. The derivative of $Q_{Gi}$ w.r.t. time can be approximated by Eqn. (13)

$$\frac{\partial Q_{Gi}}{\partial t} \approx \frac{Q_{Gi}[k+1]-Q_{Gi}[k]}{\Delta t} \tag{13}$$

Eqn. (13) can be rewritten as

$$Q_{Gi}[k+1] = Q_{Gi}[k] + \frac{\partial Q_{Gi}}{\partial t}\Delta t \tag{14}$$

where $\Delta t$ is the time interval for the control setting update.

Finally, the control variable is updated according to the designed nonlinear control law as given in Eqn. (15).

$$Q_{Gi}[k+1] = Q_{Gi}[k] - \frac{\partial f}{\partial Q_{Gi}} \Delta t \ . \tag{15}$$

The process flow for the whole formulation has been shown below in the following Fig. 1 and the corresponding pseudo code of the proposed nonlinear distributed algorithm has been explained in procedure. 1.

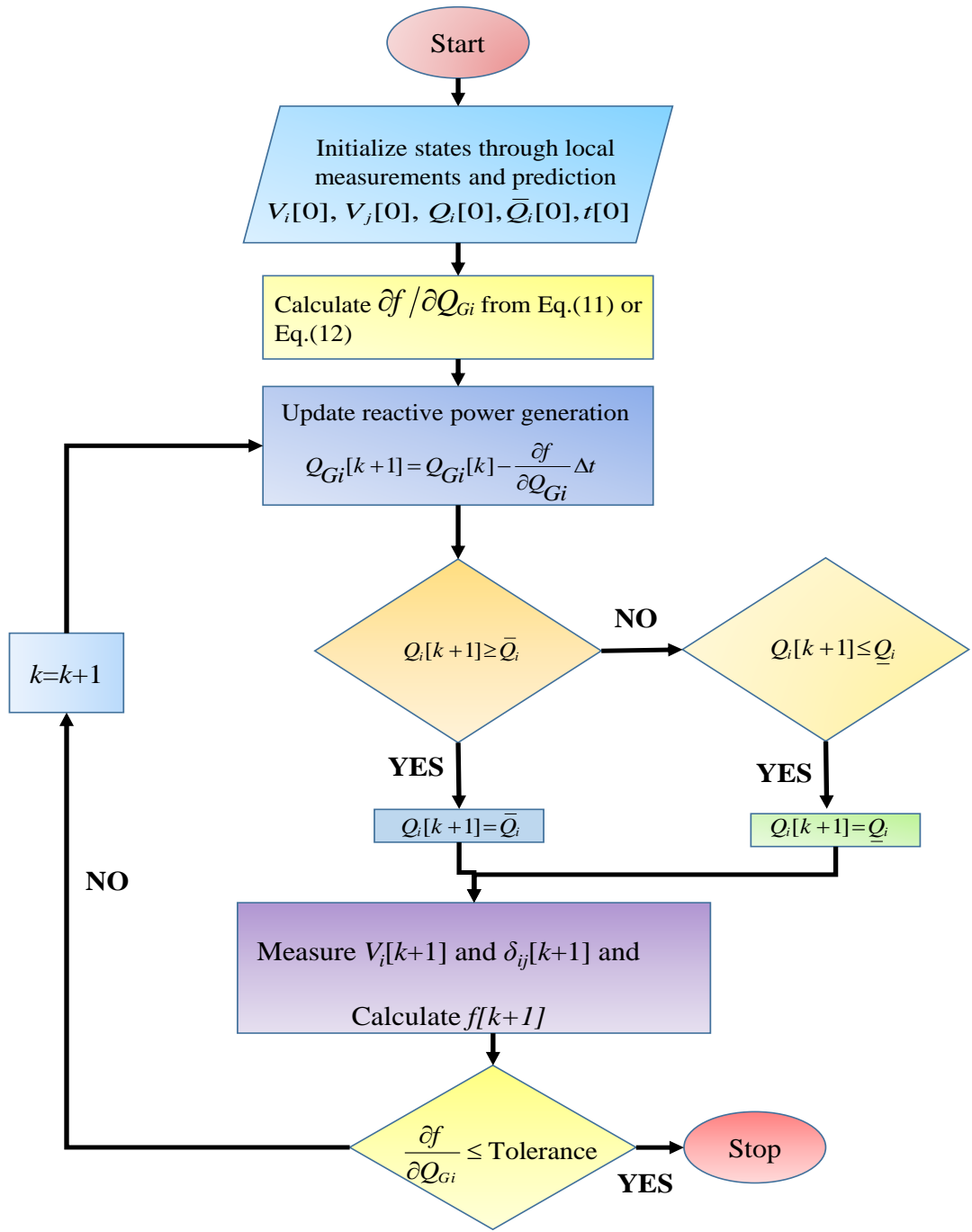

Fig.1. Flow chart to represent the algorithm

**Start**

   **Do** Initialization for the flat start

   **Do** *Step I:* Initialize $V_i[0], V_j[0], Q_i[0], \bar{Q}_i[0], t[0]$.

   **If tolerance**$<\epsilon$

   **Do** *Step II:* Calculate $\partial f/\partial Q_{Gi}$ from Eqn.(11) or (12).

   **Do** *Step III:* Update the reactive power generation by Eqn.(15).

   **Do** *Step IV:* if $Q_i[k+1] \geq \bar{Q}_i$ then $Q_i[k+1] = \bar{Q}_i$
                   if $Q_i[k+1] \leq \underline{Q}_i$ then $Q_i[k+1] = \underline{Q}_i$

   **Do** *Step V:* Measure $V_i[k+1]$, $\delta_{ij}[k+1]$ and calculate *f[k+1]*

   **Do** *Step VI:* Calculate $\dfrac{\partial f}{\partial Q_{Gi}}$ to check tolerance.

   Do *Step VII* **Else if tolerance** $>\epsilon$

   **Repeat** *Steps III-VII*

   **Else**

   **End**

Pseudo code.1 Working of the proposed distributed algorithm.

## 4. Simulation Results

In this section, to exhibit the effects of approximation of voltage angles between two neighboring buses, the proposed distributed control algorithm is applied on two different power networks: modified IEEE 9-bus system and 162-bus system [24].

### 4.1 Case Study 1: 9-Bus System

In 9-bus system, bus 1 is slack bus, bus 2 and bus 3 are the voltage controlled buses and remaining 6 buses are selected as load buses. Bus data and line data for 9 bus system is given in Table 1 and Table 2, respectively. Reactive power generators are placed at bus 5, 6, 7, 8, 9 and reactive power generation from these generators will be controlled to minimize the objective function. Cost coefficients for reactive power generation, range of reactive power, and reference voltages is shown in Table.3. Initially, weights for power loss: $w_1$, Voltage deviation: $w_2$ and reactive power cost: $w_3$ are assigned as 0.0005, 1 and 1 respectively. However, the weights can be changed to any value depending upon the preference of the reactive power suppliers.

In the first scenario, the gradient of power loss is used by approximating $\cos(\delta_{ij})$ to unity as per Eqn. (12) and then results are compared with the ones, obtained without this approximation (by including the actual value of $\cos(\delta_{ij})$ ) according to Eqn. (11).

**Table. 1**
Bus data given for 9-Bus System

| No | V(p.u) | $\delta$ (rad) | $P_G$(p.u) | $Q_G$(p.u) | $P_L$(p.u) | $Q_L$(p.u) |
|---|---|---|---|---|---|---|
| 1 | 1.04 | 0 | 0.71 | 0.27 | 0.0 | 0.0 |
| 2 | 1.025 | 9.28 | 1.63 | 0.06 | 0.0 | 0.0 |
| 3 | 1.025 | 4.64 | 0.85 | 0.10 | 0.0 | 0.0 |
| 4 | 1.025 | -2.21 | 0.0 | 0.0 | 0.0 | 0.0 |
| 5 | 0.995 | -3.98 | 0.0 | 0.0 | 1.2 | 0.5 |
| 6 | 1.012 | -3.68 | 0.0 | 0.0 | 0.9 | 0.3 |
| 7 | 1.025 | 3.71 | 0.0 | 0.00 | 0.0 | .65 |
| 8 | 1.015 | 0.72 | 0.0 | 0.0 | 0.5 | .35 |
| 9 | 1.032 | 1.96 | 0.0 | 0.0 | 0.5 | 0.7 |

**Table. 2**
Line data given for 9-Bus System

| Line From. | Line To | R(p.u) | X(p.u) | Line Charging |
|---|---|---|---|---|
| 4 | 1 | 0.00100 | 0.0576 | 0.0000 |
| 7 | 2 | 0.00500 | 0.06250 | 0.0000 |
| 9 | 3 | 0.00100 | 0.0586 | 0.0000 |
| 7 | 8 | 0.00850 | 0.07200 | 0.1490 |
| 5 | 7 | 0.01190 | 0.1008 | 0.2090 |
| 9 | 8 | 0.03200 | 0.1610 | 0.3060 |
| 9 | 6 | 0.03900 | 0.1700 | 0.3580 |
| 5 | 4 | 0.01000 | 0.08500 | 0.1760 |
| 6 | 4 | 0.01700 | 0.09200 | 0.1580 |

**Table. 3**
Summery for control parameters for 9-Bus System

| Bus No | Ap(p.u) | Bp(p.u) | Cp(p.u) | $Q_{min}$(p.u) | $Q_{max}$(p.u) | $V_{ref}$(p.u) |
|---|---|---|---|---|---|---|
| 5 | 0.082 | 2.25 | 150 | -.75 | .75 | 1 |
| 6 | 0.062 | 4.20 | 160 | -.75 | .75 | 1 |
| 7 | 0.055 | 1.25 | 140 | -.50 | .50 | 1 |
| 8 | 0.055 | 2.50 | 180 | -.75 | .75 | 1 |
| 9 | 0.053 | 2.80 | 130 | -1. | 1.0 | 1 |

To test the algorithm for real-time application, a series of load changing events are introduced in the 9 bus system as given in Table 4. Real and reactive power loads are changed on the load buses and result updates of the reactive power generation, objective function and voltage improvement are shown in Fig. (1-3) for the two cases: with and without approximation of the angle difference.

**Table 4**
Event sequences of load changes on 6-bus control area

| Event No. | Sample Time | Bus No | Load Type | Load Change |
|---|---|---|---|---|
| Event1 | 25 | 5, 6,7,8,9 | Reactive Load | 1.2×Initial |
| Event2 | 50 | 5, 6, 8, 9 | Real Load | 1.05×Event1 |
| Event3 | 75 | 5, 6,7,8,9 | Reactive Load | 0.80×Event2 |
| Event4 | 100 | 5, 6, 8, 9 | Real Load | 0.90×Event3 |

Initially, system converges to its optimal setting before the first event is introduced by increasing the reactive power load on bus no.5, 6, 7, 8, 9 at 25$^{th}$ iteration as given in Table.4. Due to increase in reactive power load demand, reactive power generation should also rise to counteract the effect of abrupt reactive power load demand change. It is important to note that proposed algorithm increases the output reactive power generation to approach to the reference voltages of the load buses immediately after the abrupt load changes take place on the energy system as shown in Figs. (1-3). It can be observed that rise in real load demand does not exhibit much rise in the reactive power generation as clear from the Event 2 results when real load is increased to 1.05 of the event1 load. Similar results are achieved when reactive power load and real power load demand decline in the Event3 and Event4, respectively.

Comparing results a) and b) for Figs. (2-4), it is clear that the convergence speed of in Fig. (a) is more than that of the (b). If Figs. (2-4) are closely analyzed, plots with approximation attain its optimal value much faster than the plots without approximation. Comparing Fig.2, before Event1, it shows that reactive power generation updates converges much faster with approximation than that of the case without approximation. It proves that voltage angle approximation reduces the computational burden and make algorithms fast.

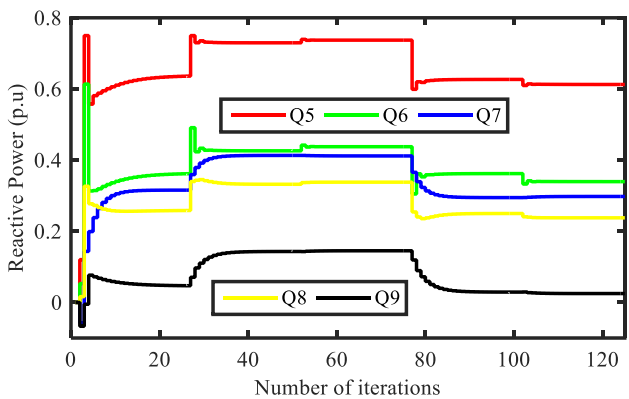 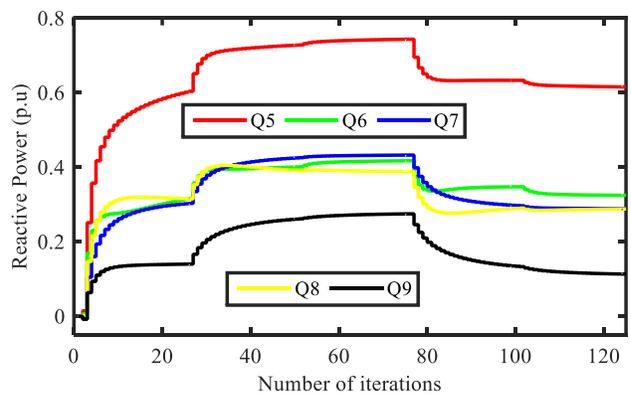

a) with approximation     b) without approximation

Fig. 2. Updates of reactive power generation

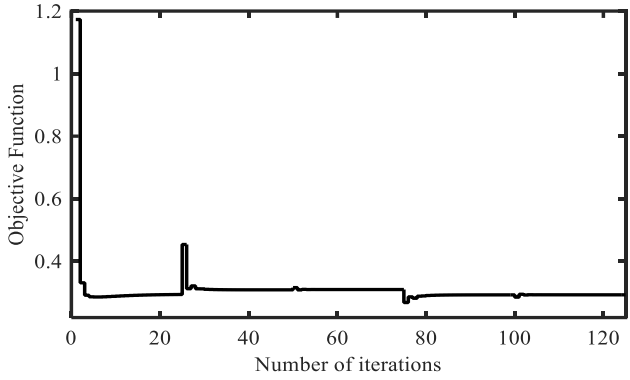 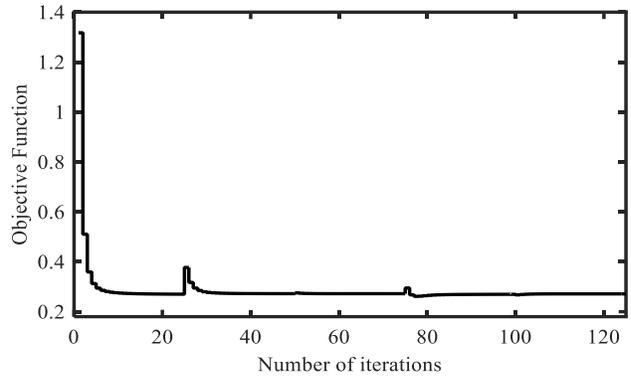

a) with approximation

b) without approximation

Fig. 3. Convergence of the objective function

To explore the impact of this approximation on optimal solution, updates of the objective functions are plotted together without introducing any series of events as shown in Fig.4. It reconfirms our assumption that with approximation, convergence speed of the algorithms can be accelerated to a great extent. However, when no approximation is performed and true value of angle difference are measured and used in the algorithm, the objective function is observed to be less than the approximated solution as shown in Fig. 5. The summary of individual cost functions for the two cases: with and without approximation is shown in Table 5. In the following sub-section, authors have attempted to explore the reasons for lower objective function value when actual value of angle difference is measured and used in the algorithm.

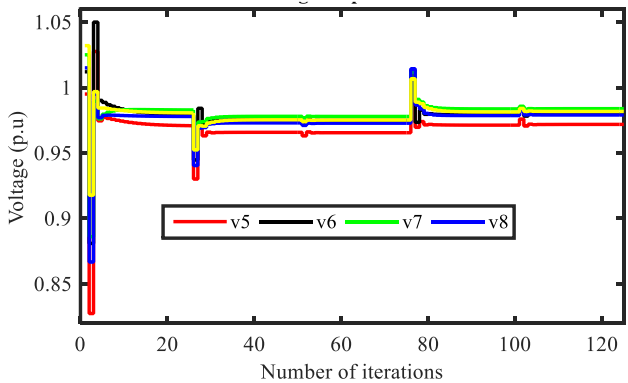 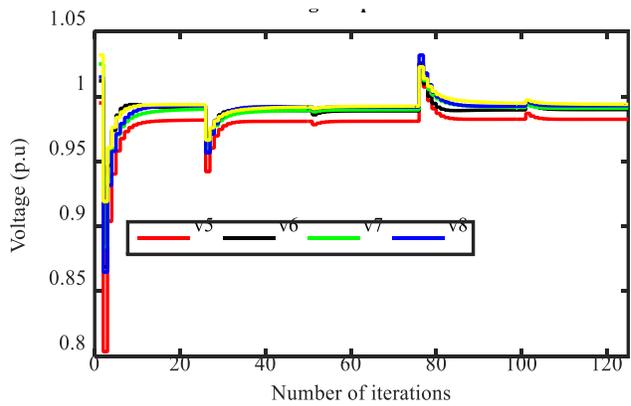

a) with approximation

b) without approximation

Fig. 4. Improvement of the load bus voltages

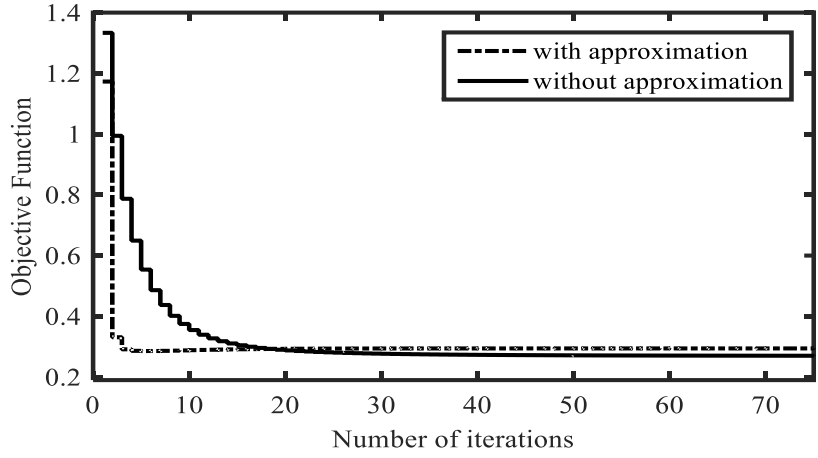

Fig. 5. Convergence of the objective function with and without approximation.

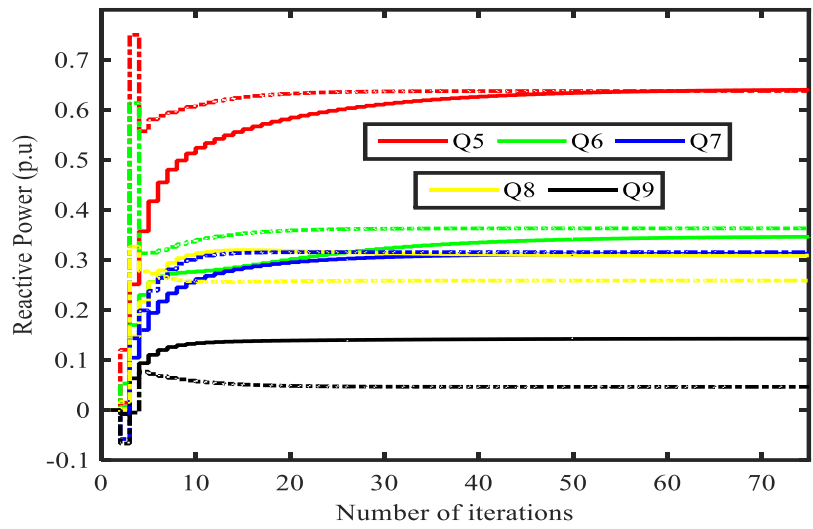

Fig.6. Reactive power generation updates for the two scenarios

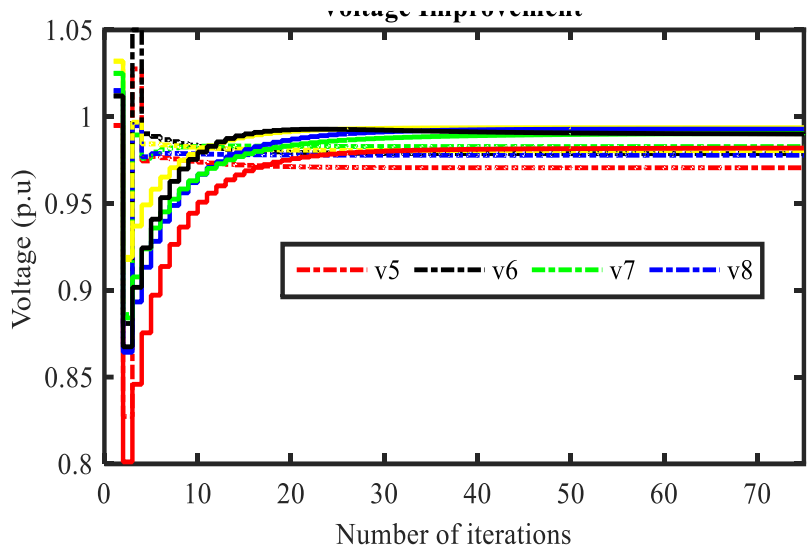

Fig. 7. Voltage profile updates with and without approximation

### 4.2 Reason for increased objective function

Table 5 shows that power loss and voltage deviation is decreased from 0.1906 p.u. and 0.0268 p.u. to 0.1869 p.u. and 0.0061 p.u. respectively, without approximation. To analyze the reason closely, reactive power generation updates and voltage profile updates for the two scenarios: with and without approximation is plotted in Fig. 6 and Fig. 7, respectively, where dotted lines show the approximated reactive power generation while solid lines show without approximation. Similarly, individual generator bus reactive power generation is summarized in Table. 6. As it can be seen that total reactive power generation is increased from 1.6216 p.u. to 1.7388 p.u. in the system, voltage deviation and power loss in the system should decrease [25-27]. The reason is that most of power systems are inductive in nature and rise in reactive power generation improves the voltage magnitude. As the voltage profile becomes close to the reference voltage, current in the line decreases because current is proportional to the voltage difference across the line. Fall in the current causes decline in power loss. Due to decline in the voltage deviation and power loss, objective function decreases as the whole. However, this change in the overall objective is not very remarkable as it is clear from Fig. 4 and Table. 5. Now authors attempts to investigate the reason of change in reactive power generation and objective function when real values of cos($\delta_{ij}$) are used in the algorithm.

**Table. 5**
Summary of individual cost functions for the 2 cases

| Bus system | Objective Function | Cost | Power Loss | Voltage Deviation |
|---|---|---|---|---|
| 9-with approximation | 0.2943 | 0.0769 | 0.1906 | 0.0268 |
| 9-without approximation | 0.2699 | 0.0769 | 0.1869 | 0.0061 |
| 162-with approximation | 3.9024 | 1.368 | 2.0965 | 0.42772 |
| 162-without approximation | 3.63807 | 1.409 | 1.8295 | 0.39957 |

### 4.3 Reason for the rise in reactive power generation

As given in Table. 6, reactive power generation for bus 5, 6 and 7 remained unchanged. However, for bus 8 and 9, it increased from 0.2588 and 0.0463 to 0.3097 and 0.1421, respectively. Reason for rise in reactive power generation for these two buses can be explained from Eqn. (16) [28-29] which states that generation is directly proportional to angle differences.

$$Q_i = Q_{Gi} - Q_{Di} = \sum_{j=1}^{n} V_i V_j (G_{ij} \sin\delta_{ij} - B_{ij} \cos\delta_{ij}) \tag{16}$$

For the first case, when cos($\delta_{ij}$) was unity, reactive power was maximum negative because sin($\delta_{ij}$).was zero. However, when real values of cos($\delta_{ij}$) are used, reactive power generation from bus $i$ will be less negative and this amount of rise in the reactive power generation depends on value of angle difference. Higher the value of angle difference, more will be the increase in the reactive power generation. It can further be explained with the help of Table 7 and Table 8 where value of cos($\delta_{ij}$) is greatly reduce to 0.9501 for line between bus 9 and 8. It is the reason why reactive power generation from 9 and 8 is increased significantly. Authors has attempted to investigate the reasons of angle differences between neighboring buses, which will be explained in the next sub-section.

Table. 6
Reactive power generation with and without approximation

| Serial No. | With | Without |
|---|---|---|
| $Q_5$ | 0.6377 | 0.6341 |
| $Q_6$ | 0.3632 | 0.3413 |
| $Q_7$ | 0.3156 | 0.3115 |
| $Q_8$ | 0.2588 | 0.3097 |
| $Q_9$ | 0.0463 | 0.1421 |
| $\sum Q_i$ | 1.6216 | 1.7388 |

Table. 7
Angle difference between all lines for 9-Bus System

| Line From | $\delta_i$(rad) | Line To | $\delta_j$(rad) | $\delta_{ij}$(rad) | $\cos(\delta_{ij})$ |
|---|---|---|---|---|---|
| 4 | -5.4970 | 1 | 00000 | -5.497 | 0.99650 |
| 7 | 8.91000 | 2 | 20.350 | -11.44 | 0.98013 |
| 9 | 0.81808 | 3 | 6.4420 | -5.624 | 0.99519 |
| 7 | 8.91000 | 8 | 2.8582 | 6.0518 | 0.99443 |
| 5 | 0.81808 | 7 | 2.8582 | -2.040 | 0.99937 |
| 9 | 8.91000 | 8 | -9.846 | 18.756 | 0.95010 |
| 9 | 0.81808 | 6 | -10.48 | 11.295 | 0.98063 |
| 5 | -9.8457 | 4 | -5.498 | -4.348 | 0.99712 |
| 6 | -10.477 | 4 | -5.498 | -4.979 | 0.99623 |

### 4.4 Reason for change of voltage angles

For detailed analysis of angle difference, Table 7 shows cosine of angle difference between neighboring buses of the connected line. It is clear that line 9-8 has the lowest value of $\cos(\delta_{ij})$=0.95010. Similarly, two more line 7-2 and 9-6 also have $\cos(\delta_{ij})$ less than 0.99 but it is greater than 0.98 and thus, can be ignored. To investigate the increase of angle difference, power loss on each line is calculated as shown in

Table. 8. Power loss on the lines exhibit the similar behavior: line 9-8 has the highest drop whereas 7-2 and 9-6 are on the second and third position respectively. It can be deduced from the available results that angle difference is directly proportional to the power loss on the line or indirectly it can be said that it is related with impedance of the line. The higher is the power loss between the lines, the more will be the angle difference of bus voltages.

**Table. 8**
Line flows and power loss for 9-Bus System

| From-To | Line flow | From-To | Line flow | Loss on line |
|---|---|---|---|---|
| 4-1 | -0.8578 | 1-4 | 0.8606 | 0.00276 |
| 7-2 | -1.5907 | 2-7 | 1.63 | 0.0393 |
| 9-3 | -0.8475 | 3-9 | 0.85 | 0.00254 |
| 7-8 | 0.67531 | 8-7 | -0.663 | 0.01211 |
| 9-8 | -0.1622 | 8-9 | 0.1632 | 0.00105 |
| 7-5 | 0.91539 | 5-7 | -0.831 | 0.08396 |
| 9-6 | 0.50962 | 6-9 | -0.476 | 0.03325 |
| 5-4 | -0.4186 | 4-5 | 0.4241 | 0.005579 |
| 6-4 | -0.4236 | 4-6 | 0.4337 | 0.01006 |

From control room of the power system, data for line flows and power loss on the lines are usually known using power flow calculation. As it is seen in the 9 bus example, majority of the lines have angle difference more than 0.98 and only few lines have low angle difference, which can be obtained from power flow data, available in the control center of the network. Thus, if power loss as % of the line flow on a line is more than 8%, angle difference approximation for those lines should not be made, however, for the rest of lines, approximation of angle difference will not have any effect on the optimal setting of the algorithms and optimal solutions may be regarded as true optimal solutions. This condition, however, will make the algorithms cost-effected and very fast.

**Table. 9**
Power loss as % of the line flow for 9-Bus System

| Line flow | From-To | Loss on line | %Loss of the line flow |
|---|---|---|---|
| -0.8578 | 1-4 | 0.00276 | 0.32064 |
| -1.5907 | 2-7 | 0.0393 | 2.411 |
| -0.8475 | 3-9 | 0.00254 | 0.29885 |
| 0.67531 | 8-7 | 0.01211 | 1.8254 |
| -0.1622 | 5-7 | 0.00105 | 0.64381 |
| 0.91539 | 9-8 | 0.08396 | 10.098 |
| 0.50962 | 6-9 | 0.03325 | 6.9807 |
| -0.4186 | 4-5 | 0.005579 | 1.3154 |
| -0.4236 | 4-6 | 0.01006 | 2.3196 |

### 4.5 Case Study 3: 162-Bus System

Effect of approximation is tested on the modified IEEE 162-bus system [24] where 16 synchronous generators are attached at various buses to control the reactive power generation as shown in Table. 10. Reference voltages are used the same as given in IEEE 162-bus data. Weight coefficient for reactive power generation cost, power loss and voltage deviation is set to be 0.0005, 1 and 1 respectively.

As it can be seen in Table. 10, reactive power generation has increased with approximation. Power loss and voltage deviation, in return, has dropped than its approximated values as shown in Table. 6. It can be analyzed from Table.10 that only one line (125-50) has % power loss/line flow value more than 10%. All the remaining lines have it less than 8%. Hence, if any AC OPF algorithm for this power network, approximation of voltage angle difference may not affect its optimal setting much, while making the algorithm faster at the same time. Thus, any AC OPF algorithm having power loss as % of the line flow less than 8%, this is a reasonable approximation. As given in the World Bank data [30], most of the developed countries have power loss less than 8%, hence, it is reasonable approximation, to be made. For developing countries, angle difference approximation is valid only for those lines having power loss less than 8%.

**Table. 10**
Reactive power generation with and without approximation for 162-bus system

| Bus No | With | Without | Difference |
|---|---|---|---|
| 3 | 2.1759 | 1.7544 | 0.04215 |
| 15 | 2.3617 | 2.3024 | 0.0593 |
| 22 | 1.646 | 1.5441 | 0.1019 |
| 27 | 2.1501 | 2.0559 | 0.0942 |
| 36 | 1.0832 | 1.0254 | 0.0578 |
| 45 | 0.54222 | 0.54972 | -0.0075 |
| 67 | 1.7714 | 1.6935 | 0.0779 |
| 68 | 1.1249 | 1.1281 | -0.0032 |
| 84 | 2.2267 | 2.2086 | 0.0181 |
| 94 | 1.7018 | 1.6339 | 0.0679 |
| 100 | 0.50493 | 0.48865 | 0.01628 |
| 124 | 0.39676 | 0.30967 | 0.08709 |
| 126 | 0.71019 | 0.68418 | 0.02601 |
| 142 | 0.36152 | 0.37924 | -0.01772 |
| 147 | 1.4772 | 1.4775 | -0.0003 |
| 148 | 0.80018 | 0.84172 | -0.04154 |
| $\sum Q_i$ | 17.4984 | 18.077 | 0.57840 |

### 4.6 Comparison of computation time

Computation time for both cases: with and with approximation are measured to estimate the effect of making such approximation. Authors used ThinkPad laptop of IBM made, with processor Intel(R) Core(TM) i7-4510U CPU @2.00Hz 2.60GHz and Installed memory (RAM) of 8GB. Operating system installed is 64 bit-operating system of Windows 10pro. The line flow and power loss for the IEEE 162 bus system and the time taken by the proposed algorithm for the two cases of 9 bus and 162 bus power system are shown in Table 11 & Table 12 respectively. It shows that as the system becomes bigger in size, the impact of the approximation becomes more evident.

In real time power system, usually the power system is quite large, with around 50,000 buses. Let us assume that we have a system of 50,000 buses and we need to perform contingency analysis within Energy Management System (EMS) to assess the possible contingency (all N-1), one at a time. If power flow takes 10 seconds to solve one contingency, it will take 7 hours to assess 2500 contingencies on a sequential computer. However, we may use parallel 10 computers to assess the contingency analysis within 42 minutes. We may further minimize the time to about 20 minutes, if we can make this approximation of voltage angle difference. This example has illustrated the importance of making such approximation in power system, which can make the algorithms computationally intelligent and fast.

**Table. 11**
Line Flows and power loss for 162-Bus System

| From Bus | To Bus | Line flow | Line Loss | Loss % of the line flow |
|---|---|---|---|---|
| 125 | 43 | 0.26529 | 0.009402 | 3.544 |
| 38 | 19 | -1.4672 | 0.052454 | 3.5751 |
| 84 | 41 | -1.2389 | 0.045798 | 3.6965 |
| 138 | 135 | -0.079443 | 0.0029815 | 3.753 |
| 74 | 26 | -7.3871 | 0.2799 | 3.7891 |
| 4 | 1 | -7.5895 | 0.28963 | 3.8162 |
| 65 | 60 | -0.4939 | 0.020631 | 4.1771 |
| 127 | 16 | -0.19149 | 0.0082229 | 4.2943 |
| 19 | 17 | -0.21654 | 0.010567 | 4.88 |
| 40 | 34 | -1.2377 | 0.061966 | 5.0065 |
| 93 | 91 | -1.6066 | 0.085219 | 5.3043 |
| 144 | 143 | 0.51361 | 0.027617 | 5.3771 |
| 21 | 17 | -0.53805 | 0.028972 | 5.3846 |
| 94 | 91 | 3.4971 | 0.19046 | 5.4462 |
| 30 | 29 | -1.3102 | 0.072937 | 5.567 |
| 126 | 16 | 0.16853 | 0.0096424 | 5.7214 |
| 115 | 111 | -1.0517 | 0.060844 | 5.7851 |
| 57 | 55 | -1.4564 | 0.085176 | 5.8484 |
| 14 | 3 | 0.36591 | 0.021654 | 5.9179 |
| 141 | 110 | -2.2403 | 0.13688 | 6.1099 |
| 77 | 34 | 3.0901 | 0.19418 | 6.284 |
| 22 | 21 | -1.2358 | 0.08022 | 6.4913 |
| 145 | 138 | -0.57848 | 0.042634 | 7.37 |
| 18 | 17 | -0.24637 | 0.018528 | 7.5204 |
| 10 | 8 | -0.20365 | 0.015322 | 7.5238 |
| 13 | 2 | 0.0039269 | 0.00023017 | 5.861366 |
| 93 | 84 | -0.80205 | 0.04765 | 5.941 |
| 21 | 19 | 0.021831 | 0.001153 | 5.2815 |
| 51 | 48 | 0.80741 | 0.050331 | 6.2336 |
| 125 | 50 | -0.67589 | 0.069633 | 10.302 |
| 146 | 144 | -0.28029 | 0.020226 | 7.21601 |
| 126 | 60 | -0.29997 | 0.015035 | 5.0121 |
| 62 | 60 | -0.016383 | 0.0011753 | 7.1739 |
| 140 | 138 | -0.02907 | 0.001607 | 5.5280 |
| 96 | 88 | -1.1547 | 0.08675 | 7.5130 |
| 52 | 48 | -0.3408 | 0.0074004 | 2.1715 |
| 146 | 142 | 0.16203 | 0.0040206 | 2.4814 |

**Table. 12**
Time comparison required for algorithms for two cases

| Bus system | With approximation | Without approximation |
|---|---|---|
| 9-Bus system | 2.04s | 2.64s |
| 162-Bus system | 3.81s | 6.64s |

### 4.7 Comparison with the modern computational intelligent tool:

To validate the effectiveness of the proposed distributed algorithm, it is compared with the modern computational intelligent tool of Particle Swarm Optimization (PSO). 30 swarm particles are used to optimize the multi-objective function where more details of PSO can be found in [7]. Results obtained by PSO are compared with our proposed distributed algorithm as shown in Table. 13. It is shown that the optimal objective function achieved using proposed distributed algorithm is approximately equal to that of intelligent tool of PSO with almost negligible error. However, our distributed algorithm is observed to be much faster than that of PSO. For 9-bus power system, proposed algorithm converges within 15 iterations whereas PSO takes 34 iterations to find the optimal solution.

**Table. 13**
Comparison of proposed algorithm with computational intelligent tool

| Bus system | Proposed distributed algorithm | | Intelligent Algorithm | |
|---|---|---|---|---|
| | Objective function | Iterations | Objective function | Iterations |
| *9-without approximation* | 0.2699 | 15 | 0.2695 | 34 |
| *162-without approximation* | 3.63807 | 25 | 3.6380 | 83 |

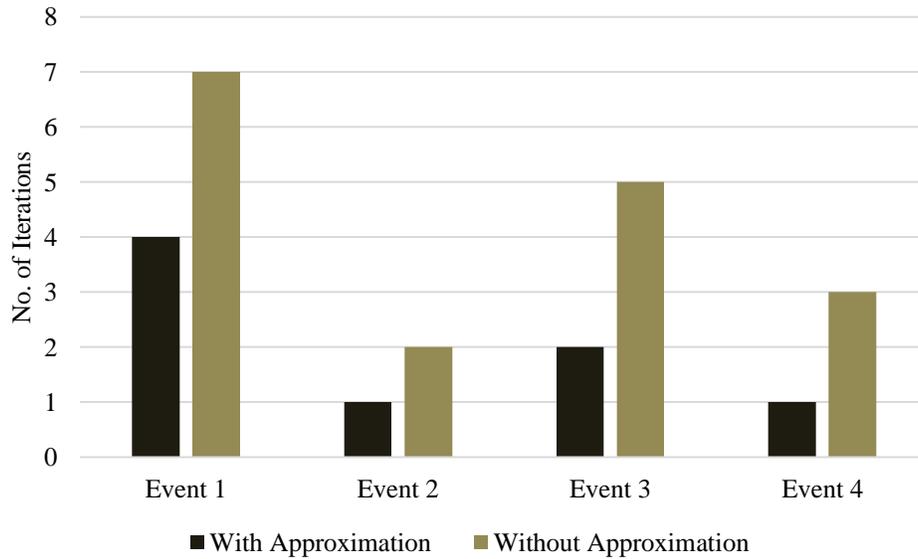

Fig.8. Settling time for line voltage for both cases after successive loading events

In order to testify for the robustness of the algorithm, the convergence rate for bus voltages to reach its new state of equilibrium after changing the loading conditions, is calculated for both the cases. It is evident from the Fig.8 that with the proposed approximation, the algorithm is more robust offering a low convergence time for both cases of reactive and active load changing conditions of Table 4.

## 5. Conclusions

This paper presented an effective way to make OPF algorithms fast. Authors have approximated voltage angle difference of the connected buses as zero and explored its impact on the computational cost, convergence speed and the optimal solutions. It is exhibited that making such approximation may lower the algorithm time significantly, with a very little impact on the optimal solutions of the algorithm. This paper also explores possible reasons of deviation of approximating solution with the true optimal solution of the algorithms and suggests a condition for making such approximation. If the power loss as % of line flow is less than 8%, it can be considered as a reasonable approximation. Finally, computation time of the proposed algorithm is compared for the two cases- namely " with" and "without" approximation.